\documentclass[prb,twocolumn,showpacs,floatfix,amsmath,amssymb]{revtex4}
\usepackage{graphicx}% Include figure files
\usepackage{dcolumn}% Align table columns on decimal point
\usepackage{bm}% bold math
\usepackage{amssymb}
\usepackage{graphicx}
\usepackage{amsmath}
\usepackage{xspace}
\begin{document}

\title{Raman phonons in $\alpha$-FeTe and Fe$_{1.03}$Se$_{0.3}$Te$_{0.7}$
single crystals}
\author{T.-L. Xia,$^{1}$ D. Hou,$^{2,1}$ S. C. Zhao,$^{1}$ A. M.
Zhang,$^{1}$ G. F. Chen,$^{3}$ J. L. Luo,$^{3}$ N. L. Wang,$^{3}$ J.
H. Wei,$^{1}$ Z.-Y. Lu,$^{1}$ and Q. M. Zhang$^{1,4}$}

\email{qmzhang@ruc.edu.cn}

\affiliation{$^{1}$Department of Physics, Renmin University of
China, Beijing 100872, P. R. China}

\affiliation{$^{2}$Department of Physics, Shandong University, Jinan
250100, P. R. China}

\affiliation{$^{3}$Beijing National Laboratory for Condensed Matter
Physics, Institute of Physics, Chinese Academy of Sciences, Beijing
100190, P. R. China}

\affiliation{$^{4}$Department of Physics, Nanjing University,
Nanjing 210093, P. R. China}

\date{\today}
\begin{abstract}
The polarized Raman scattering spectra of nonsuperconducting
$\alpha$-FeTe and of the newly discovered, As-free superconductor
Fe$_{1.03}$Se$_{0.3}$Te$_{0.7}$ are measured at room temperature on
single crystals. The phonon modes are assigned by combining symmetry
analysis with first-principles calculations. In the parent compound
$\alpha$-FeTe, the A$_{1g}$ mode of the Te atom and the B$_{1g}$ mode
of the Fe atom are observed clearly, while in superconducting
Fe$_{1.03}$Se$_{0.3}$Te$_{0.7}$, only a softened Fe B$_{1g}$ mode
can be seen. No electron-phonon coupling feature can be distinguished
in the spectra of the two samples. By contrast, the spectra of the
superconducting system show a slight enhancement below 300$cm^{-1}$,
which may be of electronic origin.
\end{abstract}

 \maketitle

The recent discovery of superconductivity in quaternary, rare-earth
transition-metal oxypnictides, and especially the subsequent raising
of the superconducting transition temperature ($T_c$) above the
MacMillan limit, has drawn great interest in the condensed matter
community.\cite{kamihara,hhwen,gfchen,xhchen,zaren}
REFeAsO$_{1-x}$F$_{x}$, which we abbreviate as FeAs-1111, is
the first series of superconductors showing such high $T_c$ values
without copper-oxide planes. As such, it provides a new system quite
different from the cuprate superconductors in which to study the
mechanism of high-temperature superconductivity. In rapid succession,
Ba(Sr,Ca)K(Na)Fe$_{2}$As$_{2}$ (FeAs-122)\cite{Rotter,gfchenFe2As2,
Sasmal,Nicrys,gfchenFe2As2crys} and Li$_{1-x}$FeAs (FeAs-111),\cite{Wang,
Pitcher,Tapp} which has an infinite layered structure, were also found
to be superconducting. It is thought that superconductivity in the
FeAs-1111 and FeAs-122 series may have a direct connection to a
spin-density-wave (SDW) anomaly occurring in the FeAs layer.\cite{cruz}
Superconductivity emerges when the SDW order is suppressed by
chemical doping or by high pressures.

All these series of iron-based superconductors contain the element
As, which is toxic on its own and would be even more so when
oxidized to As$_{2}$O$_{3}$. As a substitute, $\alpha$-FeSe with
some Se deficiency, which is less toxic and easier to handle than
arsenides, has also been found to exhibit
superconductivity.\cite{Hsu,Yeh} $\alpha$-FeSe has a PbO-type
structure, different from the NiAs-type structure which has been
studied extensively in $\beta$-FeSe.\cite{Terzieff,schuster} The
crystal structure of $\alpha$-FeSe is composed of stacked layers of
edge-sharing FeSe$_{4}$ tetrahedra, and belongs to the space group
P4/nmm. $\alpha$-FeSe has been observed to distort from tetragonal
to a triclinic structure below 105K, while its analog $\alpha$-FeTe,
which is tetragonal at room temperature, transforms to an
orthorhombic lattice at temperatures below 45K. $\alpha$-Fe(Se,Te)
has a simpler structure than the different families of Fe-based
superconductors, being essentially their infinite-layer analog.
Among the consequences of this substantial structural difference are
a shift of the SDW transition temperature by a factor of 2-3, to 65K
in Fe(Se,Te) from 140-200K in the parent compounds of the FeAs
systems. Further, the resistivity of Fe(Se,Te) shows a
semiconductor-like temperature-dependence, completely different from
the metallic behavior of the FeAs-based materials. However, it is
superconducting at temperatures up to 13 K for the composition
Fe$_{1.03}$Se$_{0.3}$Te$_{0.7}$.\cite{gfchenunp} As a safer member
of the family of iron-based superconductors, Fe(Se,Te) would seem
destined to play an important role both in fundamental research into
iron-based superconductivity and in its potential applications.

To date, the mechanism for superconductivity in the iron-based
compounds remains unclear. Many theoretical scenarios, particularly
a magnetic origin and electron-phonon coupling, have been proposed
to understand the pairing mechanism. Further, the study of the SDW
above $T_c$ is tightly correlated with the structural properties.
Thus a detailed study of the phonon modes in each series of Fe-based
superconductors can be expected to yield important clues regarding
all of the above aspects, and Raman scattering is well known to be a
unique probe of zone-center optical phonons. Raman-scattering
studies have so far been accomplished only for FeAs superconductors,
and are reported in Refs.~[\onlinecite{Hadjiev,SC Zhao,Litvinchuk}].

In this paper, we report polarized Raman-scattering results obtained
on Fe$_{1+y}$Se$_{x}$Te$_{1-x}$ single crystals, both for the $x =
0$ parent compound and for the superconducting material with $x = 0.30$.
The zone-center optical modes were classified by a group-theoretical
analysis and the Raman-active phonons assigned accordingly. We have
performed first-principles lattice-dynamics calculations using both
the relaxed and the experimental atomic positions, and find that the
latter correspond more closely to the experimental measurements.

The single crystals used in the Raman scattering experiments were
prepared by the Bridgeman technique. Details of the crystal growth
process are presented elsewhere.\cite{gfchenunp} The exact
compositions of the compounds studied here were estimated to be
FeTe$_{0.92}$ and Fe$_{1.03}$Se$_{0.30}$Te$_{0.70}$.
Raman-scattering measurements were performed with a triple Horiba
Jobin Yvon T64000 spectrometer equipped with an optical microscope
and liquid-nitrogen-cooled, back-illuminated CCD detector. A
$\times$50 long-working-distance objective was employed to focus the
laser beam, with a wave length of 532nm, into a spot of about 4$\mu$m
in diameter on the crystal surfaces, and to collect the scattered
light. The crystals were cleaved and, to avoid possible
contamination or decomposition in air, were placed immediately in a
cryostat, which was evacuated to 3$\times$10$^{-6}$Torr. The crystal
surfaces are parallel to the $ab$-plane, and Raman measurements were
conducted on these flat surfaces. Because the samples are good
metals and the crystal surfaces are perfect, most of the laser
intensity was simply reflected back. The effective incident light
intensity, and hence the Raman signal, was low, necessitating long
acquisition times.

\begin{figure}
\includegraphics[width=8 cm]{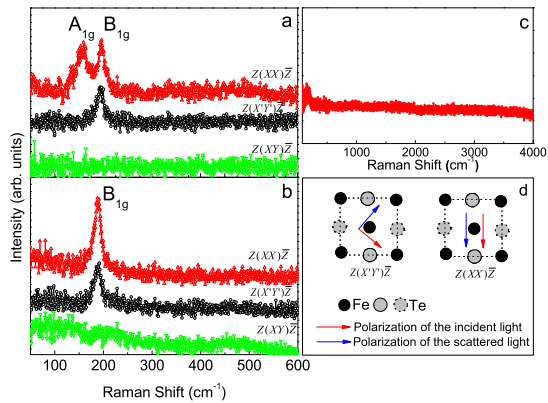}
\caption{(Color online) Room-temperature Raman spectra of FeTe$_{0.92}$
(a) and Fe$_{1.03}$Se$_{0.3}$Te$_{0.7}$ (b) for different polarized
scattering geometries in the $ab$-plane. (c) Extended spectrum of
FeTe$_{0.92}$ in polarization configuration $Z(XX)\overline{Z}$. (d)
Schematic illustration of two sample polarization configurations,
$Z(XX)\overline{Z}$ and $Z(X'Y')\overline{Z}$ (see also Fig.~2).}
\end{figure}

$\alpha$-FeTe has space group P4/nmm. The Fe and Te atoms have
respective Wyckoff positions $2a$ and $2c$. The classification of the
zone-center optical modes is similar to that of Ref.~[\onlinecite{Hadjiev}]
for SmFeAsO and LaFeAsO, and thus is not reproduced in full detail here.
By symmetry considerations, one may expect four Raman-active modes:
A$_{1g}$(Te), B$_{1g}$(Fe), E$_{g}$(Te), and E$_{g}$(Fe); the E$_{g}$
modes are two-fold degenerate. Because the experiments are performed
within the $ab$-plane, it is easy to verify that the $E_g$ modes are
absent as a consequence of their Raman tensors. Both the A$_{1g}$ and
the B$_{1g}$ mode should be present when the polarization configuration
is $Z(XX)\bar Z,$ while only the B$_{1g}$ mode should remain when the
configuration is changed to $Z(X'Y')\bar Z$. This expectation is
verified clearly by the experimental observations shown in Fig.~1(a),
where both the A$_{1g}$ and B$_{1g}$ modes of FeTe$_{0.92}$ are
found to lie lower in frequency than the equivalent modes in
Sr$_{1-x}$K$_{x}$Fe$_{2}$As$_{2}$ single crystals.\cite{Litvinchuk}
By contrast, for Fe$_{1.03}$Se$_{0.3}$Te$_{0.7}$, the A$_{1g}$ mode
of Te is absent in all scattering geometries, as shown in Fig.~1(b). To
explain this result, we note first that more impurities or vacancies may
be introduced into the Se-doped samples, because the growth conditions are
quite critical for high-quality single crystals. Secondly, the requirements
for phase formation dictate that the two samples have a small difference
in stoichiometric ratio: specifically, FeTe$_{0.92}$ corresponds to
Fe$_{1.09}$Te, while the Se-doped system is Fe$_{1.03}$Se$_{0.3}$Te$_{0.7}$.
The excess Fe ions occupy Fe(2) positions, and may thus have a substantial
effect on the vibration of neighboring Te ions. Finally, it is reasonable
to expect that 30$\%$ Te substitution by Se may in itself suppress the Te
vibration mode.

\begin{figure}
\includegraphics[width=6 cm]{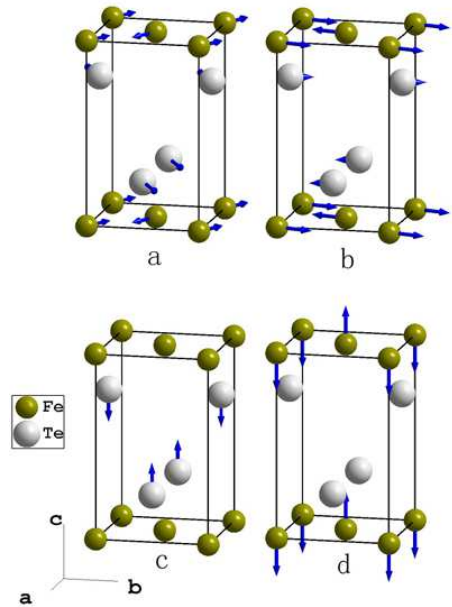}
\caption{(Color online) Displacement patterns for the Raman-active modes
of $\alpha$-FeTe, obtained from LDA calculations. (a,b) E$_{g}$ modes,
(c) A$_{1g}$ mode of Te atoms, (d) B$_{1g}$ mode of Fe atoms. }
\end{figure}

It is important to remark that a slight enhancement can be observed
in the spectra of Fe$_{1.03}$Se$_{0.3}$Te$_{0.7}$ below
300$cm^{-1}$. This may originate from electronic Raman scattering,
and further measurements are required to identify definitively the
nature of this feature. By contrast, there is no obvious feature at
all in the high-energy spectra, which were measured up to
4000$cm^{-1}$ [Fig.~1(c)]. If it were present at these energies in
the material, evidence of electron-phonon coupling would be expected
in the extended spectra.

\begin{table*}[t]
 \caption []{Left: assignment of optical phonons
in FeTe as deduced from first-principles calculations performed as
part of this study; ``NM-relaxed'' and ``NM-exp'' refer to frequencies
obtained from these calculations (see text), in units of $cm^{-1}$. Right:
experimental and relaxed cell parameters. $a$, $b$, and $c$ are the crystal
axes, while $Z$ denotes the Wyckoff positions in the $c$-direction for the
corresponding atoms.}
\begin{tabular}{|c|c|c|c|c|c||c|c|c|}
\hline
Symmetry&Atoms&NM-relaxed&NM-exp&Experiment&Active&Cell
Parameter&Relaxed&Experiment \\
\hline
$E_{g}$  & Te & 73.3 & 59.1 & & Raman & space group & P4/nmm & P4/nmm\\
$A_{1g}$ & Te & 181.1 & 140.3 & 159.1 & Raman & $a$(\AA) & 3.7078 & 3.8123\\
$E_{u}$  & Fe$+$Te & 253.8 & 195.9 & & IR & $b$(\AA) & 3.7078 & 3.8123\\
$E_{g}$  & Fe & 283.0 & 196.9 & & Raman & $c$(\AA) & 6.0326 & 6.2515\\
$B_{1g}$ & Fe & 276.2 & 215.7 & 196.3 & Raman & $Z_{Fe}$ & 0 & 0 \\
$A_{2u}$ & Fe$+$Te & 321.2 & 250.6 & & IR & $Z_{Te}$ & 0.2704 & 0.2813\\
\hline\end{tabular}
\end{table*}

To compare with the observed phonon modes, we have calculated the
non-magnetic electronic structure and the zone-center phonons of
$\alpha$-FeTe within the framework of density-functional
perturbation theory (DFPT).
%%%%%%%%%%%%%%%% LUZY makes change and add one more reference %%%%%%%%%%%%%%%%%%%%%
We applied the plane-wave basis method with the local (spin) density
approximation for the exchange-correlation potentials \cite{pw}
while the generalized gradient approximation (GGA) of
Perdew-Burke-Ernzerhof \cite{pbe} was also tested without meaningful
changes found. The ultrasoft pseudopotentials\cite{vanderbilt} were
used to model the electron-ion interactions. After the full
convergence test, the cutoffs of kinetic energy for the wave
function and for the charge-density were respectively 40 and 400
Ryd. And the gaussian smearing technique was performed on a uniform,
24$\times$24$\times$12 lattice of points in reciprocal space. We
then employed the DFPT to generate the dynamical matrix, from which
the phonon frequencies and atomic displacements were derived.
%%%%%%%%%END%%%%

The Fe$_{1+y}$Te compounds (PbO-type structure) have a narrow range
of variation in $y$.\cite{schuster} The experimental measurements
show that the excess Fe atoms partially occupy the interstitial
sites.\cite{bao} This makes first-principles calculations
complicated, because thousands of averages over sample disorder
would be required, and for this reason it was necessary to neglect
the excess Fe atoms in the phonon calculations, while the effect of
different cell parameters (experimental and relaxed) on phonons is
taken into account.

The cell parameters and calculated results are listed in Table I
together with the experimental Raman data. It is worth emphasizing
that the calculated results at zero temperature are compared to the
results of Raman experiments performed at room temperature: this is
because such a comparison ensures the same non-magnetic structures
in each case, as the ground state of Fe$_{1+y}$Te compounds at low
temperature is magnetically ordered. The temperature effect can be
included by noting that if only the temperature is increased while
the other properties of the sample (such as the magnetic state) are
unchanged, the measured phonons should soften by an amount
equivalent to the energy required for the same vibration at higher
temperatures. In this way it is possible to evaluate the effects of
temperature when assigning the phonon modes.

From Table I, the calculated phonon modes are completely consistent
with the symmetry analysis. Qualitatively, all of the Raman-active modes
were found and assigned according to the displacement patterns shown in
Fig.~2. Quantitatively, the arrows in Fig.~2 indicate not only the
vibration directions of the corresponding atoms but also, by their lengths,
represent the relative vibration amplitudes compared with those of other
atoms in the same mode. The calculated infra-red phonon frequencies are
also reported in Table I, but are not considered further here.

By comparison with the experimental Raman data, the non-magnetic phonon
calculations performed using relaxed cell parameters (denoted NM-relaxed
in Table I) failed to provide a reasonable reproduction of the B$_{1g}$
mode frequency of Fe, even when considering the effect of temperature. This
behavior is quite unlike the case in previously reported phonon calculations
on iron-based superconductors such as LaFeAsO, where the relaxed parameters
always led to more accurate results.\cite{SC Zhao} We have also verified
the calculated results using the experimental unit-cell parameters in
combination with the energy-minimized internal Te positions (not shown
here), which gives results similar to within frequency shifts of only
5cm$^{-1}$. Quite generally, relaxation of the atomic positions balances the
internal atomic forces in the unit cells, while relaxation of the unit-cell
parameters further reduces the unit-cell stresses. The consistency of the
phonon results for different ways of accomplishing this relaxation indicates
that it is not the unit-cell stresses which are responsible for the
discrepancies found when comparing with the experimental results.
However, by using the experimental cell parameters (denoted NM-exp), the
calculated B$_{1g}$ mode frequency is found to be in quite good agreement
with the experimental data. This implies that the excess of Fe atoms is
affecting the steady-state atomic positions, and further evidence for
their effects on the electronic properties has been obtained for
Fe$_{1.076}$Te.\cite{gfchenunp} This may also be one of the reasons for
the instability of the materials (below).

\begin{figure}
\includegraphics[width=6 cm]{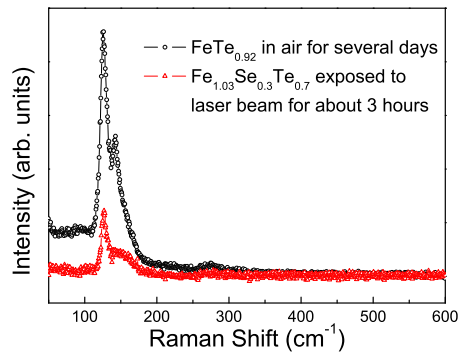}
\caption{(Color online) Room-temperature Raman-scattering spectra of
the surface from samples exposed to air for several days and from
one fixed point of the cleaved sample surface held under continuous
laser irradiation. The polarization is $Z(XX)\bar Z$.}
\end{figure}

Finally, an interesting decomposition process is observed in the
measurements. If the laser spot is held on a fixed point on the
freshly cleaved surface of Fe$_{1.03}$Se$_{0.3}$Te$_{0.7}$ in vacuum
for more than 3 hours, the measured spectrum changes completely. The
resulting spectrum is in fact quite similar to that obtained for
FeTe$_{0.92}$ after exposure to air for several days (Fig.~3). The
origin of this effect may be found in the results of
Refs.~[\onlinecite{PineTe,PineTeO2}], from which it is clear that
the anomalous spectrum originates from amorphous Te.\cite{PineTe}
Thus it is possible that Fe$_{1+y}$Se$_{x}$Te$_{1-x}$ is not stable
in air, and that one of its decomposition products is amorphous Te
rather than TeO$_{2}$. The decomposition may originate from its
rather complicated phase diagram.\cite{Yeh,ZQMao} The decomposition
of Fe$_{1+y}$Se$_{x}$Te$_{1-x}$ in air for long periods, and through
irradiation for shorter periods, may restrict the potential
applications of this sytem, and even some types of experimental
study.

In conclusion, we have measured the room-temperature, polarized
Raman-scattering spectra of FeTe$_{0.92}$ and Fe$_{1.03}$Se$_{0.3}$Te$_{0.7}$
single crystals. We performed first-principles electronic band-structure
calculations, and used these in combination with a symmetry analysis both to
identify the phonon modes of $\alpha$-FeTe and to compute their frequencies.
When Te ions are partially substituted by Se ions in superconducting
Fe$_{1.03}$Se$_{0.3}$Te$_{0.7}$, only the B$_{1g}$ mode of the Fe atom
can be observed, presumably due to the effect of random Te replacement
by Se in destroying the periodic potential of the Te atoms. The
frequency of the Fe B$_{1g}$ mode is lower in the superconductor than
in the parent compound $\alpha$-FeTe. It is noteworthy to find that
Fe$_{1+y}$Se$_{x}$Te$_{1-x}$ is not quite stable in air, one of the
decomposition products being identified as amorphous Te. By providing
detailed information on Raman phonons in FeTe-series supercondutors,
this work opens the way to further studies of the coupled structural,
magnetic, and electronic properties of these systems.

We thank Bruce Normand for fruitful discussions and for a critical reading
of the manuscript. This work was supported by the National Basic Research
Program of China (Grant Nos.~2006CB9213001 and 2006CB601002) and by the
NSFC (Grant Nos.~10574064 and 20673133).

\end{document}